\documentclass[aps,prb,showpacs]{revtex4}
\usepackage{amsmath,amssymb}
\usepackage{graphicx}
\usepackage{dcolumn}
\usepackage{bm}
\usepackage{color}
\newcommand{\mbb}{\mathbb}

\newcommand{\tet}{\texttt}
\newcommand{\mc}{\mathcal}
\newcommand{\mf}{\mathfrak}
\makeatletter
\newcommand*{\rom}[1]{\expandafter\@slowromancap\romannumeral #1@}
\makeatother

\begin{document}

\title{Exploring the Optical States for Black Phosphorus: Anisotropy and Bandgap Tuning }

\author{Andrii Iurov}
\email{aiurov@unm.edu}
\affiliation{
Center for High Technology Materials, University of New Mexico,
1313 Goddard SE, Albuquerque, NM, 87106, USA
}
\author{Liubov Zhemchuzhna}
\affiliation{Department of Physics and Astronomy, Hunter College of the City University of New York,
695 Park Avenue, New York, NY 10065, USA}
\author{Godfrey Gumbs}
\affiliation{Department of Physics and Astronomy, Hunter College of the City
University of New York, 695 Park Avenue, New York, NY 10065, USA}
\affiliation{Donostia International Physics Center (DIPC),
P de Manuel Lardizabal, 4, 20018 San Sebastian, Basque Country, Spain}
\author{Danhong Huang}
\affiliation{Air Force Research Laboratory, Space Vehicles Directorate,
Kirtland Air Force Base, NM 87117, USA}
\affiliation{Center for High Technology Materials, University of New Mexico,
1313 Goddard SE, Albuquerque, NM, 87106, USA}

\date{\today}

\begin{abstract}
The dressed states arising from the interaction between electrons and holes, and off-resonant electromagnetic
radiation have been investigated for recently fabricated gapped and anisotropic black phosphorus.  Our calculations
were carried out for the low-energy electronic subbands near the $\Gamma$ point. States for both linear 
and circular polarizations of the incoming radiation have been computed. However, our principal emphasis 
is on linearly polarized light with arbitrary polarization since this case has not been given much attention
for dressing fields imposed on anisotropic structures. We have considered various cases for 
one- and few-layer phosphorus, including massless Dirac fermions with tunable in-plane anisotropy. Initial 
Hamiltonian parameters are renormalized in a largely different way compared to those for previously 
reported for gapped Dirac structures and, most importantly, existing anisotropy which could be modified 
in every direction. 
\end{abstract}


\maketitle

\section{Introduction}
\label{sect1}

Black phosphorus (BP) is a layered structure of buckled atomic phosphorus for which the layers are 
connected by weak van der Waals forces. \cite{Jami} This is the most stable phosphorus-based crystal 
at room  temperature and for a wide range of pressure values. Crystalline BP is a semiconductor, for 
which $P$ atoms are covalently bonded with three adjacent atoms.  Such a structure exhibits strong 
anisotropy for an arbitrary number of layers, as well as hybrid electron and hole 
states in the vicinity of the band edge in phosphorene featuring both Dirac cone and a conventional 
Schr$\ddot{o}$dinger electron behavior. While bulk BP is a semiconductor with a small 
bandgap of $0.3\,eV$ , its monlayer counterpart is a semiconductor with a large direct bandgap 
$(\backsim 2\,eV)$ and is referred to as \textit{phosphorene} in analogy to graphene and could be 
exfoliated in a mechanical way.

\medskip
\par

It is not surprsing that reliable information for the band structure of BP with 
a specified number of layers has been appreciated as being extremely important for device applications,
including a field-effect transistor. \cite{Litran} This has stimulated a large number of first-principles
calculations based on the density functional theory as well as the tight-binding model \,\cite{Rudenko},
group-theoretical calculations \,\cite{Appel} and the continuum model. \cite{GGBpPRB,GGBpPRL}
 of BP structures, which recently received a significant amount of attention for their success 
in analyzing experimental data. \cite{transp1,transp2, transp3} The quality of all such 
models \,\cite{jpcmrud,japrud} depends on the accuracy of the exchange-correlation approximation. 
Over time, we  are going to witness numerous attempts to engineer new types of anisotropic electronic 
bandstruture of BP-based devices using various mechanisms such as electron-photon interaction with a 
dressing field. The corresponding effect from an imposed electrostatic field 
was addressed in Ref.~[\onlinecite{elfield}].

\medskip
\par

 The possibility of an electronic topological transition and electrically-tunable Dirac cone was 
theoretically predicted for multi-layer BP. \cite{Per17, Per18} The fact that the bandgap of BP 
is mainly determined by the number of layers was confirmed in experiment,\cite{buscema} i.e., the 
energy gap is strongly decreased for a larger number of layers and could be effectively neglected 
for $N_L > 5$. Yet demonstrating a Dirac cone with no effective mass or energy 
bandgap, such electrons still posses strong anisotropic properties for pristine black 
phosphorus. \,\cite{liklein}     Consequently, non-symmetric Klein tunneling could be 
observed \,\cite{liklein} with a shifted transmission peaks similar to electron tunneling 
for graphene in the presence of magnetic field. \cite{pe,mcc}

 \medskip
\par

Phosphorene is one of the most recently discovered members of a sizable group of low-dimensional structures 
with great potential for nanoelectronics applications. The most famous part of this family 
is graphene, fabricated in 2004. Because of its unique properties,\,\cite{gr1,gr2,gr3} graphene
has initiated a new direction in electronic devices.  A subsequent crucial advance was the discovery 
of the buckled honeycomb lattices such as silicene and germanene. Their most distinguished feature 
is the existence and tunability of two non-equivalent spin-orbit and sublattice-asymmetry bandgaps. 
The latter one, $\Delta_{z}$ is directly modified by an external 
electrostatic field. \cite{ezawaprl,ezawa9prl,ezawa, SilMain} 
This is a result of the out-of plane buckling, which is due to a larger radius 
of $Si$ and $Ge$ atoms compared to carbon and $sp^3$ hybridization. The most 
recently fabricated germanene possesses similar buckling features but with different bandgaps and
Fermi velocity. \cite{G1,G2,G3,G4}

\medskip
\par

Another important class of such materials are the \textit{transition metal dichalcogenides} 
(TMDC's) structures such as \tet{MC}$_2$, where \tet{M}  denotes a metal such as $Mo$, $W$,  
and \tet{C} is a chalcogen atom ($S$, $Se$, or $Te$). Molybdenum disulfide 
MoS$_2$, a typical representative \,\cite{habib,most} of TMDC's, has exhibited a large energy 
bandgap $\backsimeq 1.9 \,eV$, and broken symmetry between the electron and hole 
subbands so that for all experimentally accessible electron density values only 
one hole subband is doped. As a result, all the electronic, collective and transport 
properties vary significantly for the electron and hole types of doping. However, all 
these low-dimensional materials exhibit almost complete (with a slight 
deviation for MoS$_2$) isotropy in the $x-y$ plane. Consequently, phosphorene 
is a totally new material with its very unusual properties, so that complete 
and thorough studies of these characteristics open a new important chapter in 
low-dimensional science and technology. 

\begin{figure}
\centering
\includegraphics[width=0.55\textwidth]{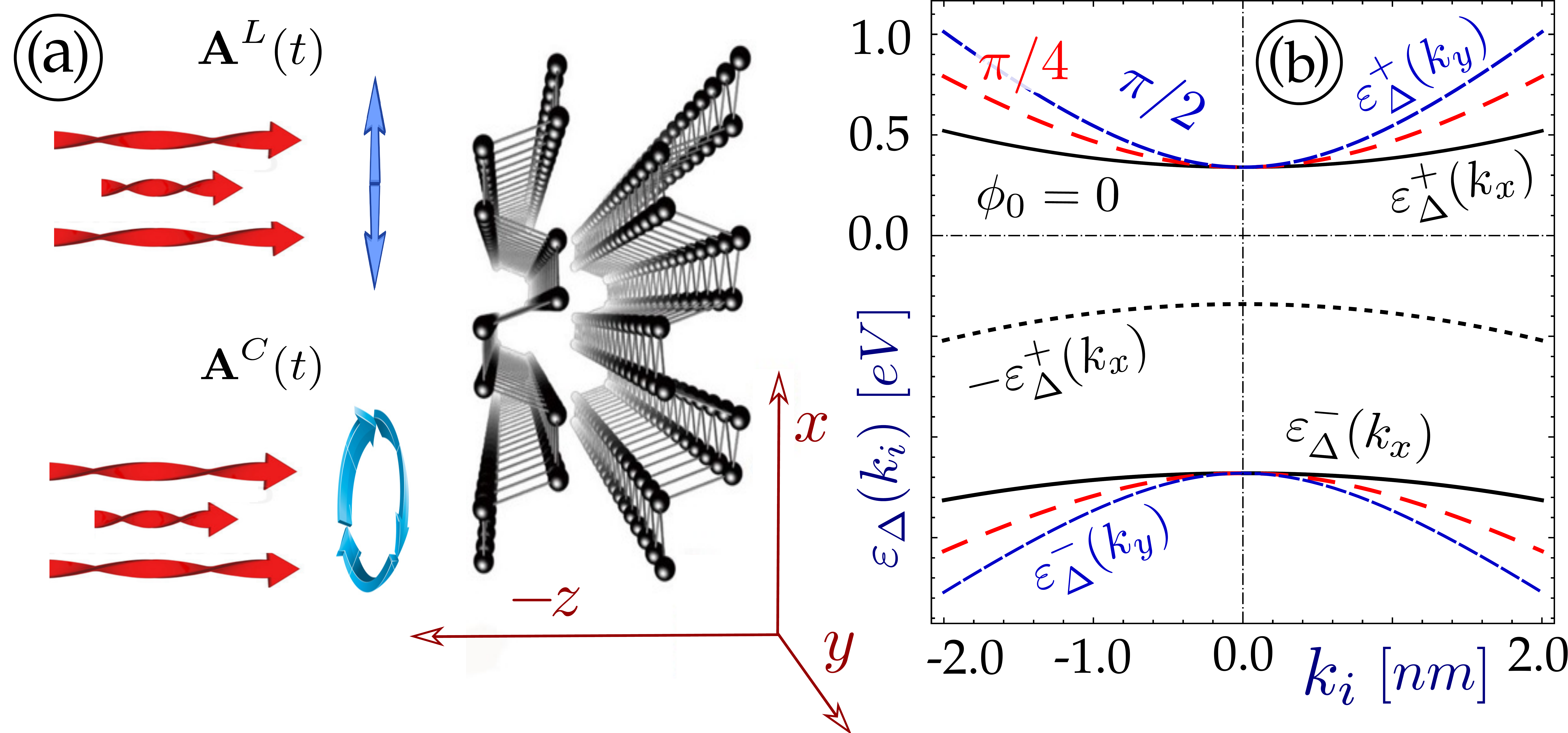}
\caption{(Color online)\
Schematics of a phosphorene layer(s) irradiated by external dressing light 
with either linear or circular polarization is shown in $(a)$. Panel $(b)$ 
presents electron and hole energy bands $\varepsilon_{\Delta}^{\pm} ({\bf k})$ 
given by Eq.\eqref{disp01} along the $x$ ($\phi_0 = 0$, solid black lines) 
and $y$ ($\phi_0 = \pi/2$, blue and short-dashed) directions, and also for the 
$(\phi_0 = \pi/4$, the red and long-dashed curve). Negative (reflected) 
\textit{electron} dispersion $- \varepsilon_{\Delta}^{+} (k_x)$ 
is also presented for comparison  to highlight the electron-hole asymmetry in phosphorene.
}
\label{FIG:1}
\end{figure}

\par
\medskip
\par
With the newest achievements in laser and microwave science, it has become possible to 
achieve substantial control and tunability of the electronic properties of low-dimensional
condensed-matter materials by subjecting them to strong off-resonant high-frequency 
periodic fields (so-called "\textit{Floquet engineering}``, schematically shown in 
Fig.~]ref{FIG:1} $(a)$).\cite{fl1,fl2,fl4,fl5} If the electron-phonon coupling 
strength is high, such a bound system could be considered as a single, holistic 
object and has been investigated using quantum optics and mechanics. These 
electrons with substantially modified energy dispersions, referred to as  
``\textit{dressed states}"', became a commonly used model in present-day
low-dimensional physics. \cite{kibis10,kibis11, kibis12,kibis19, kibis23, kibissrep}
One of the first significant achievements has been the demonstration of a
metal-insulator transition in graphene \,\cite{kibis0}, which drastically 
affected the electron tunneling and the Klein 
paradox. \cite{m1,m2} Important collective properties such as exchange and 
correlation energies are also affected by the presence of an energy gap, \cite{mPQE} 
and spin dynamics on the surface of a three-dimensional topological insulator \,\cite{kibisnjp} is also modified.

\par
\medskip
\par
The rest of our paper is organized in the following way. In Sec.~\ref{sect2}, 
we present our model, the low-energy Hamiltonian and the energy dispersions 
for phosphorene, i.e., single-layer black phosphorus with strong in-plane 
anisotropy and a large energy bandgap. The electron-photon dressed states 
for phosphorene are presented and discussed in Sec.~\ref{sect3}.  Section
  ~\ref{sect4} is devoted to calculating the dressed states for few-layer
phosphorus, in which the electrons are anisotropic massless Dirac 
fermions without a gap. Mathematical generalizations of such a model with 
both on- and off-diagonal bandgaps is considered, and the corresponding 
dressed states are also obtained. Concluding remarks are provided in Sec.~\ref{sect5}.

\section{Low-energy model for phosphorene}
\label{sect2}

Our calculations utilize the model for BP presented in Refs.~[\onlinecite{GGBpPRL, GGBpPRB}].
Being somewhat similar to the truly two-dimensional hexagon structure of graphene, the atomic 
arrangement for single layer BP results in a \textit{packered} surface due to the  $sp^3$ 
hybridization composed of the $3s$ and $3p$ orbitals. For silicene, such hybridization 
is responsible for out-of-plane ``buckling" displacement of the $Si$ atoms. 

\medskip
\par

The continuouum $k-$dependent Hamiltonian is usually based on the tight-binding model.
Close to the $\Gamma$ point, approximated up to second order in the wave vector components, 
it is given as

\begin{equation}
 \mbb{H}_{ph}^{\Delta}  ({\bf k}) = \left( 
\mbb{E}_i + \sum\limits_{i = x,y} \eta_i k_i^2
\right) \hat{\mbb{I}}_{2\times 2}+
\left( 
\Delta_O + \sum\limits_{i = x,y} \gamma_i k_i^2
\right) \hat{\Sigma}_x-
\chi k_y \hat{\Sigma}_y \, , 
\label{mosham}
\end{equation}
or in the matrix form, 

\begin{equation}
\mbb{H}_{ph}^{\Delta}  ({\bf k}) = \left[
\begin{array}{cc}
\mbb{E}_i + \sum\limits_{i = x,y} \eta_i k_i^2 & \Delta_O 
+ \sum\limits_{i = x,y} \gamma_i k_i^2 - i \chi k_y\\
\Delta_O + \sum\limits_{i = x,y} \gamma_i k_i^2 + i \chi k_y 
& \mbb{E}_i + \sum\limits_{i = x,y} \eta_i k_i^2 
\end{array}
 \right] \, .
 \label{h01}
\end{equation}
This Hamiltonian clearly displays significantly different structure and properties, 
compared to that for graphene. First, there are no linear $k-$ terms, except 
$\pm i \chi k_y$. Furthermore, there are no linear $k_x$ elements. As one of the 
most evident consequences of this structure, we note that circularly polarized 
irradiation with $x$- and $y$-components being equally important, couples to 
such electrons only in the $\backsimeq { \bf A}^2$ level.

\medskip
\par

Secondly, the energy bandgap is presented in a $\hat{\Sigma}_x$, off-diagonal 
form, contributing to the asymmetry between the electron and hole states in 
contrast to the $\hat{\Sigma}_z$-type gap. These properties, coming directly 
from the Hamiltonian structure, are new and have not been encountered previously. 
The energy dispersions are 

\begin{equation}
 \varepsilon_{\Delta}^{\pm} ({\bf k}) = \mbb{E}_i + \sum\limits_{i = x,y} \eta_i k_i^2 \pm
 \left[\left( \Delta_O  + \sum\limits_{i = x,y} \gamma_i k_i^2 \right)^2 
+ \chi^2 k_y^2  \right]^{1/2} \, ,
\label{disp01}
\end{equation}
where $\gamma_{e,h} = \pm 1$ corresponds to the electron or hole solution. For    
small values of the wave vector, these dispersions are approximated as

\begin{equation}
\varepsilon_{\Delta}^{\pm} ({\bf k}) \backsimeq \mbb{E}_i \pm \Delta_O + \left( \eta_x \pm \gamma_x \right) k_x^2 +
\left[   \eta_y \pm \left(
\gamma_y + \frac{\chi^2}{2 \Delta_O} \right)
\right] k_y^2 \ . 
\end{equation}
The effective masses given by \,\cite{GGBpPRB}

\begin{eqnarray}
&& m^{(e,h)}_{\,x} =  \frac{\hbar^2}{2 \left( \eta_x \pm \gamma_x \right)} \, , \\ 
\nonumber
&& m^{(e,h)}_{\,y} =  \frac{\hbar^2 / 2}{
\eta_y \pm \left(
\gamma_y + \chi^2 / (2 \Delta_O) \right)} \,
\end{eqnarray}
are anisotropic, and this anisotropy is different for the electron and hole 
states as $\backsim \chi^2/\Delta_O$. 

\section{Electron dressed states for in single layer}
\label{sect3}

In this Section, we calculate electron-light dressed states for phosphorene.  
As far as circularly polarized irradiation is concerned, one must consider second
order coupling in order to see how both components of the wave vector are modified.
Such consideration is critical for the vector potential given by Eq.\eqref{acirc},
but is clearly beyond the scope of conventional analytical methods. The mere presence of an 
off-diagonal energy gap $\Delta_O$ means that there is no electron/hole symmetric solution,
obtained in Refs.~[\onlinecite{kibissrep,kibisall}]. This situation also leads us to 
conclude that the Hamiltonian parameters, such as the energy gap, are affected at 
lower order than such parameters for Dirac fermions. Consequently, for   monolayer
BP, we focus on the case for linearly polarized irradiation.

\subsection{Linear polarization of the dressing field and induced anisotropy}

Since electron energy dispersion relations and their effective masses are intrinsically
anisotropic for phosphorenes, the direction of the dressing field polarization 
now plays a substantial role. We define this direction by an arbitrary angle 
$\theta_0$ from the $x-$axis and generalize the vector potential used in 
Ref.~[\onlinecite{kibisall}] so that it now has both $x-$ and $y-$ non-zero components 

\begin{equation}
 {\bf A}^{L}(t) =  \left\{\frac{E_0}{\omega} \cos \, \omega t\cos \theta_0; \,\frac{E_0}{\omega} \cos \, \omega t \sin \theta_0 \right\} \, .
 \label{Alin}
\end{equation}

\medskip
\par

The renormalized Hamiltonian for the dressed states is obtained by the canonical substituion 
$k_{x,y} \Rightarrow k_{x,y} - e /\hbar A_{x,y}$,  where $e$ stands for the electron charge, yielding

\begin{equation}
 \hat{\mc{H}}(k) = \mbb{H}_{ph}^{\Delta} + \hat{h}_{0} + \hat{h}_{int} \, , 
 \label{th}
\end{equation}
where $\mbb{H}_{ph}^{\Delta}$ is the ``bare", non-interacting electron Hamilotonian, 
given by Eq.\eqref{h01}. The zeroth-order, $k-$independent 
\textit{interaction Hamiltonian} may be expressed as

\begin{equation}
\hat{h}_{o} = \chi \, \frac{E_0 e}{\hbar \omega} \, \sin \theta_0 \, \cos \omega t \, \hat{\Sigma}_y = 
c_0 \, \left(
\begin{array}{cc}
0  & -i \cos \omega t\\
i \cos \omega t &  0
\end{array}\right) \ ,
\end{equation}
where $c_0 = \chi \,\sin \theta_0  \, E_0 \mf{e}/(\hbar \omega) $, $\mf{e} = \vert e \vert$. 
We conclude that the vector potentail of linearly polarized light \eqref{Alin}  
must have a non-zero $y-$ component in order to enable $\backsim {\bf A}^L$ coupling. 
We now turn our attention to the interaction term which is linear in $k_{x,y}$ and given by

\begin{equation}
 \hat{h}_{int} =
 2 \frac{e}{\hbar} \left( 
 \sum\limits_{i = x,y} \eta_i k_i \, A_i^{L}(t) \, \mbb{I}_{2 \times 2} + 
 \sum\limits_{i = x,y} \gamma_i k_i \, A_i^{L}(t) \, \hat{\Sigma}_x
 \right) \, ,
 \end{equation}
 or, introducing the following simplifying notations 
$\epsilon_{\alpha} = \sqrt{\alpha_x^2 k_x^2 + \alpha_y^2 k_y^2}$, 
 $\phi^{(\alpha)} = \tan^{-1} \left[ \alpha_y k_y / (\alpha_x k_x) \right]$ 
for $\alpha = \eta, \gamma$ and 
$c^{(2)} = (2 \mf{e} E_0) / (\hbar \omega)$, we express it as

\begin{equation}
\hat{h}_{int} = c^{(2)} \, \cos \, \omega t \,
\left[\begin{array}{cc}
 \epsilon_{\eta} \cos (\phi^{(\eta)} - \theta_0) & 
 \epsilon_{\gamma} \cos (\phi^{(\gamma)} - \theta_0)   \\
 \epsilon_{\gamma} \cos (\phi^{(\gamma)} - \theta_0) & 
 \epsilon_{\eta} \cos (\phi^{(\eta)} - \theta_0)  
 \end{array}
 \right] \, .
\end{equation}
As the fisrt step, we solve a time-dependent Schr\"odinger equation 
for ${\bf k} = 0$:

\begin{equation}
 i \hbar \frac{\partial \Psi_0(t)}{\partial t} = \hat{h}_{o} \Psi_0(t) \, . 
\end{equation}
We obtain the solution in a  straightforward way to be

\begin{equation}
\Psi_0^{\beta = \pm 1}(t) = \frac{1}{\sqrt{2}} \,
 \left[
 \begin{array}{c}
 1 \,\\
\beta \, i
 \end{array}
 \right] \tet{exp}\left\{
 - i \beta \,  \frac{c_0}{\hbar \omega} \, \sin \omega t
 \right\} \, .
 \label{str}
 \end{equation}
It is noteworthy that even if both energy bandgaps $\Delta_O$ and the initial 
energy shift $\mbb{E}_i $ are included so that the Hamiltonian takes the form

\begin{equation}
 h_{o} = \mbb{E}_i \, \mbb{I}_{2 \times 2} + \Delta_O \, \hat{\Sigma}_x + c_0 \, \cos \omega t \, \hat{\Sigma}_y = 
 c_0 \, \left(
 \begin{array}{cc}
\mbb{E}_i & \Delta_O -i \cos \omega t\\
\Delta_O + i \cos \omega t &  \mbb{E}_i
\end{array}
\right) \, ,
\end{equation}
the solution could still could be determined analytically as 

\begin{equation}
\Psi_0^{\beta = \pm 1}(t) = \frac{1}{\sqrt{2}} \,
 \left[
 \begin{array}{c}
 1 \\
\beta \, i
 \end{array}
 \right] \tet{exp}\left\{
 - \beta \frac{i}{\hbar} \left[
 \frac{c_0}{\omega} \, \sin \omega t - (\Delta_O - \beta \mbb{E}_i) \, t
 \right]
 \right\} \, . 
 \end{equation}
We note that such a trivial solution could only be obtained for equal diagonal energies 
$\mbb{E}_0$, i.e., for an introduced non-$\Sigma_z$ type of energy gap. If the diagonal energies
are not equivalent, given by $\mbb{E}_{1,2} = \mbb{E}_i \pm \Delta_D$, complete symmetry between 
the two components of the wave function no longer exists and such a solution could 
not be determined analytically. Consequently, we will use basic set in 
Eq.\eqref{str} for the rest of the present calculation. Such a situation (non-zero diagonal 
gap $\Delta_D$) appears if there is a relatively small non-zero vertical electrostatic 
field component and the symmetry between the vertically displaced phosphorus atoms
is broken similar to silicene. We will often use this mathematical generalization in the rest
of our work.

\par
\medskip
\par

For finite wave vector, we present the eigenfunction as an expansion in terms of
a basis set \,\cite{kibisall}

\begin{equation}
 \Psi_{\bf k} (t) = \mc{F}^{\Uparrow}\, \Psi_0^{+}(t) + \mc{F}^{\Downarrow} \, \Psi_0^{-}(t) \, ,
 \label{15}
\end{equation}
where $\mc{F}^{\Uparrow,\Downarrow} = \mc{F}^{\Uparrow,\Downarrow}(k_x,k_y \, 
\vert \, t)$ are scalar, time-dependent coefficients with anisotropic $k$-dependence. 
This equation immediately results in the two following indentities

\begin{equation}
 i \hbar \frac{d \, \mc{F}^{\, \Uparrow,\Downarrow}}{dt} = 
 \langle \Psi_0^{\pm}(t) \vert \delta\mbb{H}_{ph}^{\gamma}({\bf k}) \vert \Psi_0^{+}(t) \rangle \,
 \mc{F}^{\, \Uparrow} + \langle \Psi_0^{\pm}(t) 
\vert\delta\mbb{H}_{ph}^{\gamma}({\bf k}) \vert \Psi_0^{-}(t) \rangle \,
\mc{F}^{\, \Downarrow} \, ,
\label{16}
\end{equation}
where 

\begin{equation}
\delta\mbb{H}_{ph}^{\gamma}({\bf k}) = \mbb{H}_{ph}^{\gamma} + \hat{h}_{int} 
\end{equation}
is the bandgap and wave vector dependent portion of the total Hamiltonian \eqref{th}. 
This system becomes

\begin{eqnarray}
 i \frac{d \, \mc{F}^{\, \Uparrow,\Downarrow}}{dt} && = 
\left[
\mbb{E}_i + \eta_x k_x^2 + ( \eta_y k_y \pm \chi ) k_y + c^{(2)} 
\epsilon_{\eta} \cos (\phi^{(\eta)} - \theta_0 ) \, \cos{\omega t}
\right] 
\, \mc{F}^{\, \Uparrow, \Downarrow} \pm \\
\nonumber
&& \pm \,\, 
i \left[
\mp i \Delta_D + \Delta_O + \gamma_x k_x^2 + \gamma_y k_y^2 
+ c^{(2)} \epsilon_{\gamma} \cos (\phi^{(\gamma)} - \theta_0 ) \, \cos{\omega t}
\right]
\, \mc{F}^{\, \Downarrow, \Uparrow} 
\, \tet{exp}\left[
\pm 2  i  \,
\frac{c_0}{\hbar \omega} \, .
\, \sin \omega t \right] \,
\end{eqnarray}

\par
\medskip
\par
The quisiparticle energy dispersion relations $\varepsilon_d ({\bf k})$ 
are calculated by using the Floquet theorem from the following substitution

\begin{equation}
\mc{F}^{\Uparrow, \Downarrow}(t) = \tet{exp}\left[ - i \frac{\varepsilon_d ({\bf k}) }{\hbar} \, t \right] 
\sum\limits_{\lambda = - \infty}^{\infty} f_\lambda^{\Uparrow, \Downarrow} \, 
 \tet{e}^{i \lambda \omega t} \, ,
\end{equation}
where $f^{\Uparrow, \Downarrow}(t) = f^{\Uparrow, \Downarrow}(t + 2 \pi/ \omega)$ are 
time-dependent periodic functions. The nested exponential dependence is usually 
simplified using the Jacobi-Anger identity

\begin{equation}
\tet{exp}\left[
\pm 2  i  \,
\frac{c_0}{\hbar \omega} \,
\, \sin \omega t 
\right] = 
\sum\limits_{\nu = - \infty}^{\infty} \tet{e}^{i \nu \omega t} \, J_\nu \left(
\frac{\pm 2 c_0}{\hbar \omega}\right)\ .
\end{equation}
The orthonormality of Fourier expansions results in the following system of 
$2 \mu$, $\mu \Rightarrow \infty$ equations

\begin{eqnarray}
\nonumber
&&  \varepsilon_d ({\bf k}) \, f^{\Uparrow, \Downarrow}_\mu = 
\left[
\mu \, \hbar \omega + \mbb{E}_i \pm \eta_x k_x^2 + ( \eta_y k_y \pm \chi ) k_y
\right] \,f^{\Uparrow, \Downarrow}_\mu +     \left[ \frac{c^{(2)}}{2} 
\epsilon_{\eta} \cos (\phi^{(\eta)} -\theta_0) \right]
\,  \left( f^{\Uparrow, \Downarrow}_{\mu+1} + f^{\Uparrow, \Downarrow}_{\mu-1} \right) 
\sum\limits_{\lambda = -\infty}^{\infty}
 f^{\Downarrow, \Uparrow}_\lambda \times \\ 
 && \times  \, \left\{  \left[ \Delta_D 
 \pm  i  \left( \Delta_O + \gamma_x k_x^2 + \gamma_y k_y^2 \right)
 \right]  \, J_{\mu-\lambda} \left( \frac{\pm 2 c_0}{\hbar \omega}
 \right) + \left[ \frac{c^{(2)}}{2} \epsilon_{\gamma} 
\cos (\phi^{(\gamma)} - \theta_0) \right] \,  \sum\limits_{\alpha = \pm 1}
 J_{\mu+ \alpha -\lambda} \left( \frac{\pm 2 c_0}{\hbar \omega} \right) \right\}  \, .
\label{main}
\end{eqnarray}  
In our consideration, the frequency of the off-resonant dressing field is high 
enough so that only diagonal elements in the eigenvalue equation \eqref{main} are 
retained. However, if we need to include the first-order electron-field coupling terms
$\backsim c^{(2)}$, we must keep the summations with $\lambda = \mu \pm 1$. 

\medskip
\par 

\begin{figure}
\centering
\includegraphics[width=0.49\textwidth]{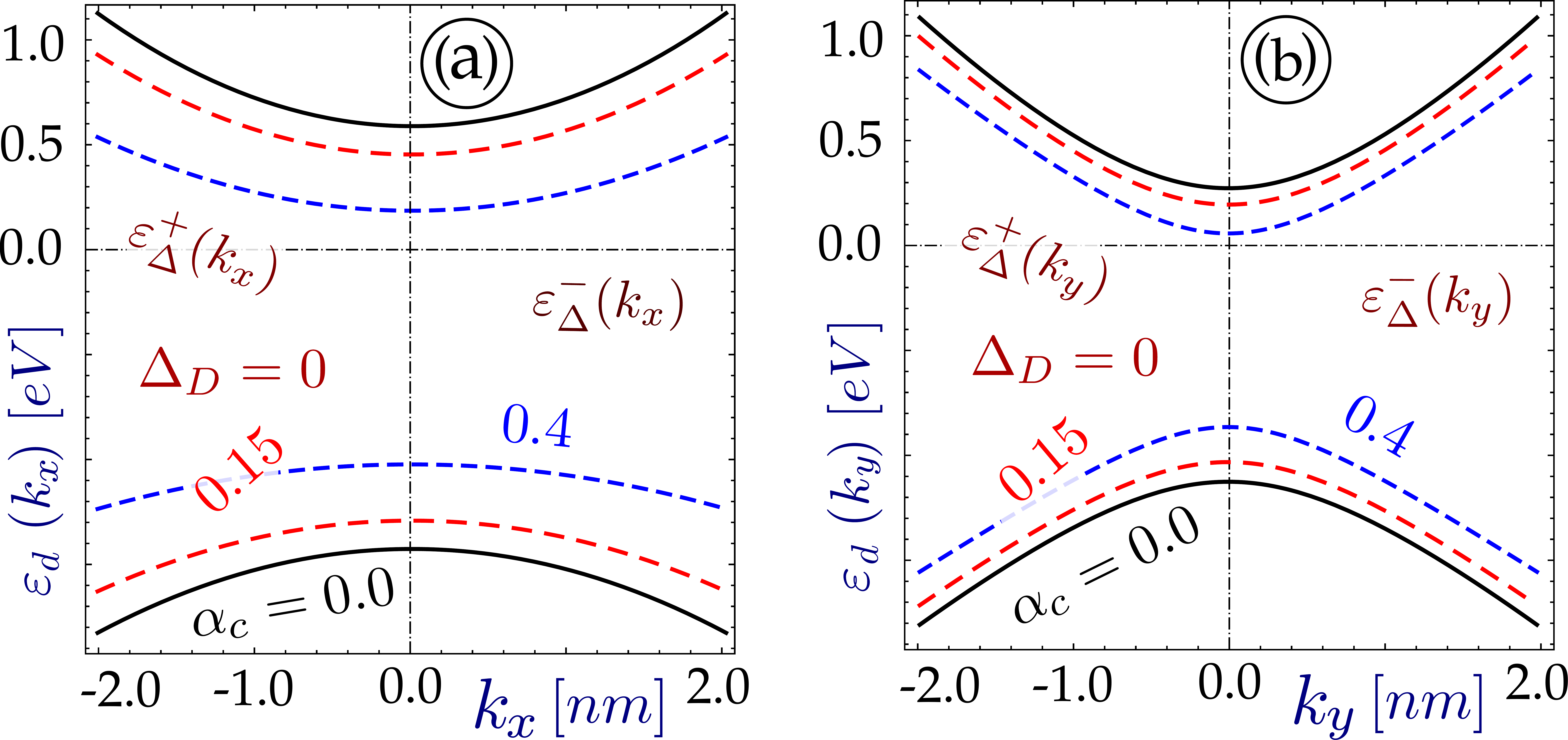}
\caption{(Color online)\ Eenrgy dispersions of electron dressed states in 
phosphorene interacting with linearly-polarized light. The two panels correspond
to the $x-$ and $y-$components of the wave vector. For each plot, the chosen dimensional 
electron-photon coupling constant is $\alpha_c = 0.0$ (black solid line), 
$\alpha_c = 0.15$, red and dashed, and $\alpha_c = 0.4$ (blue small-dashed curve).}
\label{FIG:2}
\end{figure}

In the simplest case, where only diagonal elements are kept, the quasiparticle 
energy dispersion relations are    
 
\begin{equation}
\varepsilon_d ({\bf k}) = \mbb{E}_i + \sum\limits_{i=x,y} \eta_i k_i^2 \pm \left\{
\chi^2 k_y^2 + \left[\Delta_D^2 + \left(\Delta_O + \sum\limits_{j=x,y} 
\gamma_j k_j^2 \right)^2 \right] \, J_0^2  \left(
\frac{2 c_0}{\hbar \omega} \right) \right\}^{1/2} \, .
\end{equation}
This result is valid only if $c_0 = \chi \,\sin \theta_0  \, E_0 \mf{e}/(\hbar \omega) 
\neq 0$, or there is a finite $y-$component of the polarization direction 
of the dressing field.

\begin{equation}
 \varepsilon_d ({\bf k}) = \mbb{E}_i \pm \left\{
 \left(1 - \alpha_c^2\right) \left[
 \Delta_D^2 + \Delta_O^2
 \right]
\right\}^{1/2} + \left[
\eta_x \pm  \gamma_x \, \frac{\sqrt{1 - \alpha_c^2} \, \Delta_O}{
\sqrt{
 \Delta_D^2 + \Delta_O^2}} \right] \, k_x^2 + \left[
\eta_y \pm \frac{\gamma_y \, \Delta_O \, \left(1 - \alpha_c^2\right) + \chi^2/2}{ \sqrt{
\left(1 - \alpha_c^2 \right) \left[ \Delta_D^2 + \Delta_O^2 \right]}}\right]\, k_y^2 \, , 
\end{equation}
where $\alpha_c  = 2 c_0 / (\hbar \omega)$ is a dimensionless coupling coefficient. 
The electron effective masses are now readily obtained. If there is no $\Sigma_z$-type 
energy bandgap $\Delta_D$, the expressions  are simplified as

\begin{eqnarray}
&& m^{(e,h)}_{\,x} =  \frac{\hbar^2}{2 \left( \eta_x \pm \tilde{\alpha}_c \, 
\gamma_x \right)} \, , \\ 
\nonumber
&& m^{(e,h)}_{\,y} =  \frac{\hbar^2 / 2}{
\eta_y \pm \left[ \tilde{\alpha}_c \,
\gamma_y + \chi^2 / (2 \Delta_O \tilde{\alpha}_c \,) \right]} \, ,
\end{eqnarray}
where $ \tilde{\alpha}_c = \sqrt{1 - \alpha_c^2} \backsim 1 - \alpha_c^2 / 2$. 
The obtained energy dispersion relations are presented in Fig.~\ref{FIG:2}. 
It is interesting to note that both energy bandgaps are renormalized by the 
electron-photon interaction, showing a substantial decrease, while the diagoanl 
terms for the initial effective masses of a ``bare" electron $\eta_{x,y}$
are unchanged. 

\section{Anisotropic massless fermions in few-layer phosphorus}
\label{sect4}

\par
The central property, as well as the research focus of phosphorene, is the electron 
dispersion relation and effective mass anisotropy. At the same time, BP-based materials
have a band gap which is determined by the number of layers, which varies from 
$0.6\,eV$ for five layers to $1.5\,eV$ for a single layer. Specifically, we consider 
anisotropic massless Dirac particles, which could be observed in special few-layer 
($N_L > 5$) black phosphorus superlattices for a narrow range of energies.

\medskip
\par

This anisotropic Dirac Hamiltonian 

\begin{equation}
\mbb{H}_{ml}^{\gamma_0} = \hbar v_F' \left( k_x \hat{\Sigma}_x  + \gamma_0 k_y \, 
\hat{\Sigma}_y  \right)
\end{equation}
leads to the following energy dispersion

\begin{equation}
 \varepsilon_{\gamma_0}^{\pm} ({\bf k}) = \pm \hbar v_F' \sqrt{k_x^2 + (\gamma_0 k_y)^2} \, .
\end{equation}
For such fermions interacting with light having linear polarization in arbitrary 
direction $\theta_0$ described by Eq.\eqref{Alin}, we obtain the following Hamiltonian

\begin{equation}
\hat{\mc{H}}({\bf k}) =  \mbb{H}_{ml}^{\gamma_0}({\bf k}) + \hat{h}_0 \, ,
\end{equation}
where

\begin{equation}
 \hat{h}_0 = \mf{e} v_F  \, \mathbf{\hat{\Sigma}} \cdot {\bf A}^{L}(t) = \frac{\mf{e} v_F E_0}{\omega} \,
 \left(
 \begin{array}{cc}
  0  & \tet{e}^{ - i \theta_\gamma} \cos  \omega t \\
\tet{e}^{ i \theta_\gamma} \cos  \omega t &  0
\end{array} \right)
\label{h00}
\end{equation}
is the ${\bf k} = 0$ portion of the total Hamiltonian. Here, we also introduced 
$\theta_\gamma = \tan^{-1}\left[\gamma_0 \tan(\theta_0)\right]$ so that 
$\tet{e}^{ \pm i \theta_\gamma} = \cos \theta_0 \pm i \gamma_0 \sin \theta_0$. 
We define $c_0 = \mf{e} v_F E_0 / \omega$ to be the electron-photon
interaction coefficient with the dimension of energy and for a given range of frequency 
$c_0 \ll \hbar \omega$  - dressing field which cannot be absorbed by electrons.
Traditionally, we first need to solve the time-dependent Schr\"odinger equation 
for ${\bf k} = 0$ and Hamiltonian $\hat{h}_0$. The eigenfunction is obtained in a 
straightforward way as

\begin{equation}
\Psi_0^{\beta = \pm 1}(t) = \frac{1}{\sqrt{2}} \,
 \left[
 \begin{array}{c}
 1 \,\\
\beta \tet{e}^{i \theta_\gamma}
 \end{array}
 \right] \tet{exp}\left\{
 - i \beta \,  \frac{c_0}{\hbar \omega} \, \sin \omega t
 \right\} \, .
 \label{setafm}
 \end{equation}

\medskip
\par 
In order to determine the solution for a finite wave vector we once again employ
the expansion \eqref{15} and solve Eq.~\eqref{16} for the time-dependent coefficients
$\mc{F}^{\Uparrow,\Downarrow} = \mc{F}^{\Uparrow,\Downarrow}(k_x,k_y \, \vert \, t)$. 
In our case, this leads to  
    
\begin{equation}
  i \frac{d \, \mc{F}^{\, \Uparrow,\Downarrow}}{dt} = \pm v_F' k_\gamma 
	\cos(\phi^{(\gamma)} - \theta_\gamma) \, \mc{F}^{\, \Uparrow,\Downarrow} 
  \pm i \, v_F'k_\gamma \sin(\phi^{(\gamma)} - \theta_\gamma) \, \mc{F}^{\, 
	\Downarrow,\Uparrow} \, \tet{exp}\left[\pm 2  i  \,
\frac{c_0}{\hbar \omega} \,
 \, \sin \omega t 
 \right] \, ,
\end{equation}
where $\phi^{(\gamma)} = \tan^{-1}[\gamma_0 k_y/k_x]$ or $\phi^{(\gamma)} 
= \tan^{-1}[\gamma_0 \tan(\phi_0)]$ and $\tan \phi_0 = k_y/k_x$. We also denote
$k_\gamma = \sqrt{k_x^2 + (\gamma_0 k_y)^2}$.

\par
\medskip
\par
Now we also use the Floquet theorem to extract the quasiparticle energy  $\varepsilon_d ({\bf k})$ 
and expand the remaining time-periodic functions as a Fourier series. The result is 
again a system of $2 \mu$, $\mu \Rightarrow \infty$ equations

\begin{eqnarray}
 \varepsilon_d ({\bf k}) f^{\Uparrow, \Downarrow}_\mu = \sum\limits_{\lambda = -\infty}^{\infty} 
 \left\{
  \delta_{\mu, \lambda} \, \left[
  \mu \, \omega \pm \hbar v_F' k_\gamma \cos (\phi^{(\gamma)} - \theta_\gamma)
  \right] \,f^{\Uparrow, \Downarrow}_\lambda 
 \pm  i  \hbar v_F' k_\gamma \sin (\phi^{(\gamma)} - \theta_\gamma)  \, J_{\mu-\lambda} \left(
\frac{\pm 2 c_0}{\hbar \omega}
 \right) \, f^{\Downarrow, \Uparrow}_\lambda
 \right\} \, .
\end{eqnarray}
\begin{figure}
\centering
\includegraphics[width=0.49\textwidth]{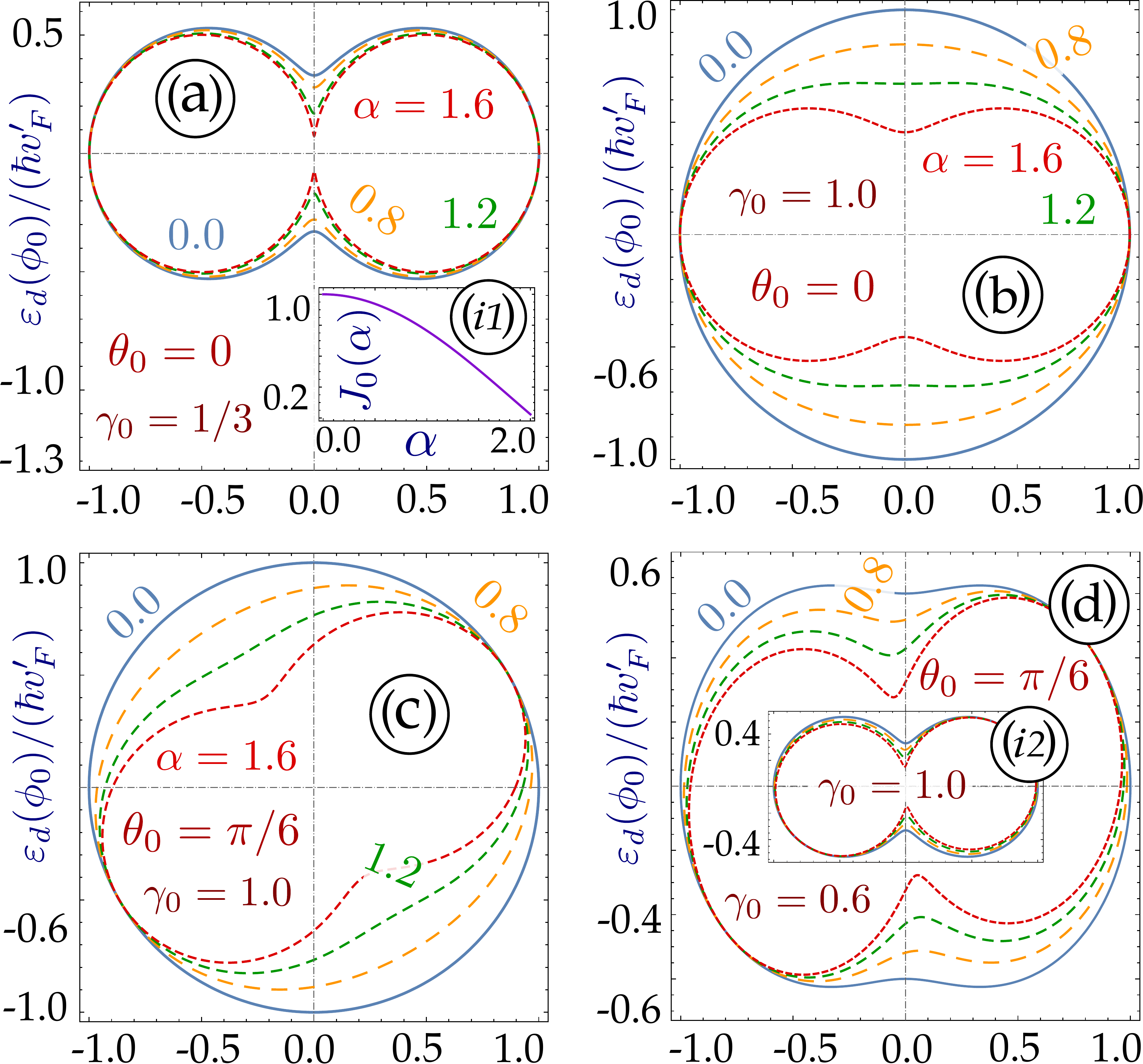}
\caption{(Color online)\ Angular dependence of the energy dispersion for anisotropic 
mass fermions (AMF's) subjected to the linearly polarized dressing field for chosen 
$\vert {\bf k} \vert = 1.0\,nm$, shown as polar plots. Each panel corresponds to   
different polarization directions $\theta_0$ and Dirac cone anisotropy parameter 
$\gamma_0$. Plot $(a)$ shows the case when $\theta_0 = 0$ (x-axis linear polarization) 
and $\gamma_0 = 1/3$, while the inset $(i1)$ demonstrates the way in which the electron-field 
interation affects the dispersion relations as a zeroth-order Bessel function. 
Panels $(b)$ and $(c)$ represent angular dispersion for $\gamma_0 = 1$ (isotropic Dirac cone) 
and $\theta_0 = 0$ and $\pi/6$, respectively. Panel $(d)$ describes the situation for 
$\theta_0 = \pi/6$ and $\gamma_0 = 0.6$ for the main plot, and $\gamma_0 = 0.3$ in inset $(i2)$. 
For each panel, the electron-photon coupling parameter $\alpha = 2 c_0 /(\hbar \omega) 
= 0.0$ (no irradiation), $0.8$, $1.2$ and $1.6$. }
\label{FIG:3} 
\end{figure}

\par
In the region of interest, i.e., for large frequency $\omega \gg v_F' k$ and 
$\omega \gg \epsilon ({\bf k})$ we approximate  
$f^{\Uparrow, \Downarrow}_{\mu \neq 0}  \backsimeq 0$. Finally, the eigenvalue 
equation becomes $\varepsilon_d ({\bf k}) f^{\Uparrow, \Downarrow}_0 = \tensor{K}(\Uparrow, \Downarrow \, \vert \, \gamma_0, {\bf k}) 
\times { \bf f}^{\Uparrow, \Downarrow}_0$, where

\begin{equation}
 \tensor{K}(\Uparrow, \Downarrow \, \vert \, \gamma_0, {\bf k}) = \left[
 \begin{array}{cc}
  \hbar v_F' k_\gamma \cos (\phi^{(\gamma)} - \theta_\gamma) & i \gamma_0 \, \hbar v_F' k_\gamma \sin (\phi^{(\gamma)} - \theta_\gamma)  
  \, J_0 \left[
2 c_0/(\hbar \omega)
 \right] \\
  - i \hbar v_F' k_\gamma \sin (\phi^{(\gamma)} - \theta_\gamma)  \, J_0 \left[
- 2 c_0/(\hbar \omega)
 \right] & - \hbar v_F' k_\gamma \cos (\phi^{(\gamma)} - \theta_\gamma)
 \end{array}
 \right]
\end{equation}
The energy eigenvalues are given by

\begin{equation}
 \varepsilon_d ({\bf k}) = \pm \hbar v_f' \sqrt{k_x^2 + (\gamma_0 k_y)^2} \, \left\{
 \cos^2 (\phi^{(\gamma)} - \theta_\gamma) +\left[ \sin (\phi^{(\gamma)} - \theta_\gamma) \, J_0 \left(
\frac{2c_0}{\hbar \omega}
 \right) \right]^2   
 \right\}^{1/2} \ .
\end{equation}
For small light intensity $c_0 \ll \hbar \omega$ the zeroth-order Bessel function of
the first kind behaves as

\begin{equation}
J_0^{c_0 \ll \hbar \omega} \left[
\frac{2c_0}{\hbar \omega}
\right] \backsimeq 1 - \frac{c_0^2}{(\hbar \omega)^2} + \frac{c_0^4}{4 (\hbar \omega)^4} - ...
\end{equation}
and we have approximately for the energy dispersion 

\begin{equation}
\varepsilon_d ({\bf k}) = \pm \hbar v_f' \left\{
\left[1 - \frac{2c_0^2}{(\hbar \omega)^2} \sin^2 \theta_\gamma \right]^2 k_x^2 + \gamma_0^2 \, 
\left[1 - \frac{2c_0^2}{(\hbar \omega)^2} \cos^2 \theta_\gamma \right]^2 k_y^2 + \frac{2\gamma_0 c_0^2}{(\hbar \omega)^2} k_x k_y
\sin (2 \theta_\gamma) 
\right\}^{1/2} \, .
\end{equation}
If the light polarization is directed along the $x$-axis, then $\theta_0 = \theta_{\gamma} = 0$ 
and 

\begin{equation}
\varepsilon_d ({\bf k}) = \hbar v_f' \sqrt{
 k_x^2 + \gamma_0^2 \, 
\left[1 - \frac{2 c_0^2}{(\hbar \omega)^2} \right]^2 k_y^2 } \, .
\end{equation}
Angular dependence of the dressed states dispersions is shown in Fig.~\ref{FIG:3}.
We notice that initially existing in-plane anisotropy is affected for all in-plance 
angles, depending on the dressing field polarization direction. For small intensity 
of the incoming radiation, polarized along the $x-$ axis, the anisotropy 
coefficient is simply renormalized. 

\subsection{Circular Polarization}

\medskip
\par

For circular polarization of the dressing radiation, the vector potential is 

\begin{equation}
 {\bf A}^C(t) = \{ A_{0,x}, \, A_{0,y} \} = \frac{E_0}{\omega} \{ \cos \, 
\omega t, \,\, \sin \, \omega t \} \, .
 \label{acirc}
\end{equation}
Being completely isotropic, this type of field is known to induce the metal-insulatron 
transition in graphene,\cite{kibis0} resulting in the creation of a non-zero energy bandgap. 
If such a gap already exists, it could be increased or decreased depending
on its initial value. \cite{kibisall} At the same time, the slope of the Dirac dispersions, 
known as Fermi velocity and the in-plane isotropy are not changed. The situation cannot be 
the same for an initially anisotropic Dirac cone for AMF's.

\medskip
\par
The total Hamiltonian for the interacting quasiparticle now becomes

\begin{equation}
 \hat{\mc{H}}({\bf k}) =  \mbb{H}_{ml}^{\gamma_0}({\bf k}) + \hat{h}_0^{(c)} \, ,
\end{equation}
where the ${\bf k} = 0$ interaction term is

\begin{equation}
\hat{h}_0^{(c)} = \frac{e v_F E_0}{\omega} \, \mathbf{\hat{\Sigma}} \cdot {\bf A}^C (t) = 
 \frac{c_0}{2} \, \left[
 \begin{array}{cc}
  0  & \sum\limits_{\alpha = \pm 1} (1 - \alpha \gamma) \, \tet{e}^{i\alpha \gamma \omega t} \\
  \sum\limits_{\alpha = \pm 1} (1 + \alpha \gamma) \, \tet{e}^{i\alpha \gamma \omega t} &  0
 \end{array}  \right] \ .
\end{equation}
It seems rather surprising, although physically justified, that this problem is 
mathematically identical to the isotrpic Dirac cone interacting with elliptically 
polarized light addressed in Refs.~[\onlinecite{kibisall, goldprx}]. The interaction 
term also could be presented as 

\begin{eqnarray}
 && \hat{h}_0^{(c)} = \hat{\mbb{S}}_\gamma \, \tet{e}^{i \omega t} 
+ \hat{\mbb{S}}_\gamma^{\dagger} \, \tet{e}^{ -i \omega t} \\
 \nonumber
 &&  \hat{\mbb{S}}_\gamma = \frac{c_0}{2}  \sum\limits_{\alpha = \pm 1} (1 + \alpha 
\gamma) \hat{\Sigma}_\alpha \, ,
\end{eqnarray}
where $\hat{\Sigma}_{\pm} = 1/2 \left(\hat{\Sigma}_x \pm i \hat{\Sigma}_y \right)$.
This Hamiltonian represents an example of a wide class a periodically driven quantum 
systems. \cite{goldprx} Such problems are generally solved perturbatively, in powers of 
$c_0/(\hbar \omega)$, if the electron-field coupling is weak. The effective Hamiltonian 
for such problem has been shown to be
 
\begin{equation}
 \hat{\mc{H}}_{eff}({\bf k}) \backsimeq  \mbb{H}_{ml}^{\gamma_0}({\bf k}) + 1 / (\hbar \omega) \, \left[ \, 
 \hat{\mbb{S}}_\gamma\,  \hat{\mbb{S}}^\dagger_\gamma \,
 \right] +  \frac{1}{2 (\hbar \omega)^2} \, \left\{ \, \left[ \left[ \, 
 \hat{\mbb{S}}_\gamma, \mbb{H}_{ml}^{\gamma_0}({\bf k}) \, \right] \, \hat{\mbb{S}}^\dagger_\gamma \, 
 \right] +  h.c. \right\} + \cdots
\end{equation}
Evaluating this expression, we obtain 
\begin{equation}
  \hat{\mc{H}}_{eff}({\bf k}) =   \hbar v_F' k_x \left(
  1 -\frac{\gamma_0 \, c_0^2}{2 (\hbar \omega)^2} 
  \right) \, \hat{\Sigma}_x + \hbar v_F' \, \gamma_0 k_y \left(
  1 -\frac{c_0^2}{2 (\gamma_0 \, \hbar \omega)^2} 
  \right) \, \hat{\Sigma}_y  -\frac{c_0^2}{\hbar \omega} \,\gamma_0 \, \hat{\Sigma}_z \, .
\end{equation}
Finally, the energy dispersion is given by
   
\begin{equation}
  \varepsilon_d ({\bf k}) = \pm \left\{
 \left(\frac{\gamma_0 \, c_0^2}{\hbar \omega} \right)^2 + \hbar^2 v_F'^2 \left[
 \left(
  1 -\frac{\gamma_0 \, c_0^2}{2 (\hbar \omega)^2} 
  \right)^2 k_x^2 + \gamma_0^2 \left(
  1 -\frac{c_0^2}{2 (\gamma_0 \, \hbar \omega)^2} 
  \right)^2 k_y^2 
 \right]   \right\}^{1/2} \, .
\end{equation}
This result is an approximation. As we mentioned, in the case of an isotropic 
Dirac cone, electrons in graphene interacting with a circularly-polarized dressing 
field, the energy gap was found to be \,\cite{kibis0,kibissrep}

\begin{equation}
 \Delta_g/2 = \sqrt{\hbar^2 \omega^2 + 2 \, c_0^2 } - \hbar \omega 
 \backsimeq \frac{c_0}{\hbar \omega} - \frac{1}{4} \left( \frac{c_0}{\hbar \omega} \right)^2 
 + ... \,\,\, ,
\end{equation}
while the Fermi velocity $v_F$ is unaffected.

\subsection{Gapped anisotropic fermions}

\medskip
\par
We now present a generalization of previously considered massless Dirac particles 
with a finite energy bandgap. Two different gaps added to both on- ($\Delta_D$) and 
off-diagonal terms ($\Delta_O$) of the Hamiltonian. Here, an anisotropic Dirac cone 
is combined with the energy gaps attributed to phosphorene, a single-layer structure.  
Even though this model does not exactly describe any of the fabricated black phosphorus 
structures, we conider it as an interesting mathematical generalization of the anisotropic 
Dirac fermions case, which may become relevant from a physical point of view. 
Apart from that, this is an intermediate case between phosphorene and few-layer 
gapless materials, which is expected to approximate the electronic properties of 
a system with a small number of phospohrus layers.  We have 

\begin{eqnarray}
\mbb{H}_g =  &&  \mbb{E}_i \, \hat{\mbb{I}}_{2\times 2} +
\left( \hbar v_F' k_x + \Delta_O \right) \hat{\Sigma_x}  + \hbar v_F' k_y  \hat{\Sigma}_y + \Delta_D 
\hat{\Sigma}_z   = \\
\nonumber
 = &&  \left[
\begin{array}{cc}
\mbb{E}_i + \Delta_D & 0 \\
0 &  \mbb{E}_i - \Delta_D
\end{array}
\right]  + \Delta_O
\left[  \begin{array}{cc}
  0 & 1 \\
  1 &  0
 \end{array}  \right]  +
 \hbar v_F' \left[
 \begin{array}{cc}
  0 &  k_x - i \gamma_0 k_y   \\
 k_x + i \gamma_0 k_y  & 0
 \end{array}
 \right] \, .
\end{eqnarray}
The corresponding energy dispersion is given by 

\begin{equation}
\epsilon_{\gamma_0}^{\pm}(k) = \mbb{E}_0 \pm \left\{
\Delta_D^2 + \left[\Delta_O + \hbar v_F' k_x \right]^2 + (\gamma_0 \, \hbar v_F k_y)^2
\right\}^{1/2} \ .
\end{equation}
We now address the interaction of these gapped Dirac electrons with the dressing 
linearly polarized field. The vector potential here
is again specified by Eq.\eqref{Alin} and the new Hamiltonian is

\begin{equation}
 \hat{\mc{H}}({\bf k}) =  \mbb{H}_{g}({\bf k}) + \hat{h}_0 \, ,
\end{equation}
where $\hat{h}_0$ is identical to Eq.\eqref{h00} since both equations share 
similar $k-$dependent terms. Following this approach, we expand the wave function 
for a finite wave vector $\Psi_{\bf k} (t)$ over the basis \eqref{setafm} and obtain 
the following equations for the expansion coefficients 
$\mc{F}^{\, \Uparrow,\Downarrow} (k_x, k_y \, \vert \, t)$

\begin{eqnarray}
  i \dot{\mc{F}}^{\, \Uparrow,\Downarrow} && = \pm \left\{
  \pm \mbb{E}_i + (\hbar v_F' k_x + \Delta_O)\cos \theta_\gamma + \gamma_0 \, \hbar v_F' k_y \sin \theta_\gamma
  \right\}
  \, \mc{F}^{\, \Uparrow,\Downarrow}  + \\
  \nonumber
&& + i \left\{
- i \Delta_D \pm 
\gamma_0 \, \hbar v_F' k_y \cos \theta_\gamma  \mp (\hbar v_F' k_x + \Delta_O) \sin \theta_\gamma
\right\}
\, \mc{F}^{\, \Downarrow,\Uparrow} \, \tet{exp}\left[ \pm 2 i \,
\frac{c_0}{\hbar \omega} \,
 \, \sin \omega t 
 \right] \, . 
\end{eqnarray}
Similar to our previous case, we introduce the following simplifying notation 

\begin{eqnarray}
 && \phi^{(O)} = \tan^{-1} \left\{ \frac{
 \gamma_0 \, \hbar v_F' k_y
 }{
 \hbar v_F' k_x + \Delta_O
 } \right\} \\
 \nonumber
 && \epsilon_{O} = \sqrt{(\hbar v_F' k_x + \Delta_O)^2 + (\gamma_0 \, \hbar v_F' k_y)^2} \ .
\end{eqnarray}
After performing the Floquet theorem substitution and the expansions similar
to the previously adopted procedure and we obtain

\begin{eqnarray}
 \varepsilon_d ({\bf k}) f^{\Uparrow, \Downarrow}_\mu =   && 
  \left[
  \mu \, \omega + \mbb{E}_i \pm \epsilon_{O} \cos (\phi^{(O)} - \theta_\gamma)
  \right] \,f^{\Uparrow, \Downarrow}_\lambda + \\
  \nonumber  
  + && \sum\limits_{\lambda = -\infty}^{\infty} \left[ \Delta_D 
 \pm  i  \epsilon_{O} \sin (\phi^{(O)} - \theta_\gamma) 
 \right]
 \, J_{\mu-\lambda} \left(
\frac{\pm 2 c_0}{\hbar \omega}
 \right) \, f^{\Downarrow, \Uparrow}_\lambda
 \, .
\end{eqnarray}
It is interesting to note that the diagonal bandgap $\Delta_D$ is now on the 
main diagonal and is affected by the Bessel function. The energy dispersions are now
given by

\begin{eqnarray}
 \varepsilon_d ({\bf k}) && = \mbb{E}_i \pm \left\{
 \left[\epsilon_{O} \, \cos (\phi^{(\gamma)} - \theta_\gamma)\right]^2 +\left[ \Delta_D^2 +
  \epsilon_{O}^2 \, \sin^2 (\phi^{(\gamma)} - \theta_\gamma)\right] \, J_0^2 \left(
\frac{2c_0}{\hbar \omega}
 \right)    
 \right\}^{1/2} \, , 
 \label{1p}
 \\
 \nonumber
 \epsilon_{O} && = \sqrt{(\hbar v_F' k_x + \Delta_O)^2 + (\gamma_0 \, \hbar v_F' k_y)^2} \,\, .
\end{eqnarray}
If the electron-photon interaction is small, then $\alpha_c = 2 c_0/(\hbar \omega) \ll 1$,
the energy dispersion relation is approximated by

\begin{eqnarray}
\label{2p}
\left( \varepsilon_d ({\bf k}) - \mbb{E}_i \right)^2  && \backsimeq (1 - \alpha_c^2) \, \Delta_D^2 +
\left[1 - \alpha_c^2 \sin^2 \theta_\gamma \right]^2 (\hbar v_F' k_x + \Delta_O)^2 + (\hbar v_F')^2 \, \gamma_0^2 \, 
\left[1 - \alpha_c^2 \cos^2 \theta_\gamma \right]^2 k_y^2 + \\
\nonumber
&& + \,\, 2 \gamma_0 \, \alpha \, \hbar v_F' (\hbar v_F' k_x + \Delta_O)
k_y \, \sin (2 \theta_\gamma) \, .
\end{eqnarray}
From Eqs.~\eqref{1p} and \eqref{2p}  we note that the diagonal $\Delta_D$ bandgap 
is decreased as $\backsim \alpha_c^2$, similar to that for gapped graphene or the 
transition metal dichalcogenides, \cite{kibisall} while the off-diagonal gap 
modification drastically depends on the direction of the radiation polarization. 
This behavior has no analogy in the previously considered structures. At the same time, 
the Fermi velocity components are modified similarly to that for anisotropic massless 
fermions. 

\section{Concluding Remarks}
\label{sect5}

\medskip
\par

We have derived closed-form analytic expressions for electron-photon dressed states in one- and few layer 
black phosphorus. The energy gap is determined by the number of layers comprising the system, reaching its largest
value for a single layer (phosphrene) and effectively vanishes for a few-layer structure. The latter case gives rise to   
anisotropic massless fermions, which exhibits an anisotropic Dirac cone. Since the anisotropy is the most significant 
and common property. Of all cases of BP, we focused on linearly polarized dressing field in an arbitrary direction.
As a result, we demonstrated the Hamiltonian parameters are modified in an essentially different way compared to
all isotropic Dirac systems. Anisotropy of the energy dispersion is modified in all directions, as well as all the 
electron effective masses.   If both diagonal and off-diagonal gaps are present, the latter one remains unaffected, 
but only for a specific light polarization direction. For AMF's interacting with circularly polarized light, 
the problem is mathematically identical with Dirac electrons irradiated by the field with elliptical polarization. 
In that case we found that initially absent energy bandgap is created and the non-equivalent Fermi velocities 
in various directions are renormalized. These results are expected to be of high importance for electronic 
device applications based on recently discovered and fabricated black phosphorus. 

\medskip
. 
\acknowledgments
D.H. would like to thank the support from the Air Force Office of Scientific Research (AFOSR).

\bibliography{DSPh}

\end{document}